# Distortion Search – A Web Search Privacy Heuristic

Kato Mivule
Department of Computer Science
Norfolk State University
Norfolk, Virginia, USA

Kenneth Hopkinson
Department of Electrical and Computer Engineering
Air Force Institute of Technology
Dayton, Ohio, USA

*Abstract*—Search engines have vast technical capabilities to retain Internet search logs for each user and thus present major privacy vulnerabilities to both individuals and organizations in revealing user intent. Additionally, many of the web search privacy enhancing tools available today require that the user trusts a third party, which make confidentiality of user intent even more challenging. The user is left at the mercy of the third party without the control over his or her own privacy. In this article, we suggest a user-centric heuristic, Distortion Search, a web search query privacy methodology that works by the formation of obfuscated search queries via the permutation of query keyword categories, and by strategically applying k-anonymised web navigational clicks on URLs and Ads to generate a distorted user profile and thus providing specific user intent and query confidentiality. We provide empirical results via the evaluation of distorted web search queries in terms of retrieved search results and the resulting web ads from search engines. Preliminary experimental results indicate that web search query and specific user intent privacy might be achievable from the user side without the involvement of the search engine or other third parties.

*Keywords—Web search privacy; query obfuscation; user profile privacy; user intent obfuscation*

I. INTRODUCTION

Special thanks to the US Air Force Institute of Technology (AFIT) and US National Research Council (NRC) for supporting this work.

Search engines have enormous technological capabilities in maintaining user Internet web search logs and as such, present privacy vulnerabilities to both individuals and organizations whereby user intent could be revealed without user consent. Of recent the issue of web search privacy got national attention when the US Congress overturned Internet Privacy Regulations instituted by the previous Obama administration that prevented search engines and Internet service providers (ISP) from selling web search and browsing history of their clients at will [1]. To add to the excitement, many media outlets published articles on how to circumvent search engines and ISPs by suggesting that users utilize randomness and noise in their web searches and browsing history [2]-[5]. It is at this same time that these events have found us working on the Distortion Search Hueristic, a contribution, we believe, might help users control their own personal privacy. Additionally, this situation is further compounded by the fact that many of the web search privacy enhancing tools available today require that the user trusts a third party, making the confidentiality of user intent even more challenging. In such cases the user is left at the mercy of the third party and without control over his or her own privacy. In this article, we propose a user-centric confidentiality search heuristic, Distortion Search, a web search query privacy technique that works by the formation of obfuscated search queries via the permutation of query keyword categories, and by strategically applying k-anonymised web navigational clicks on URLs and Ads to generate a distorted user profile and thus providing specific user intent and query confidentiality. We provide empirical results via the evaluation of distorted web search queries in terms of retrieved search results and the resulting web ads from search engines. Preliminary experimental results indicate that web search query and specific user intent privacy might be achievable from the user side without the involvement of the search engine or other third parties. The rest of the article is organized as follows. In Section II, background and related works are discussed. In Section III, the Distortion Search Heuristic methodology is considered. Preliminary results from the experiment are discussed in Section IV. Finally, the conclusion future works is given in Section V.

II. BACKGROUND AND RELATED WORK

*A. Background and Terms*

*Data privacy* is the protection of an entity's information from disclosure without the explicit consent of that entity. User intent in the context of web search queries on the Internet could be considered as confidential information [6]. *A web search engine* is a software application used to search for information on the internet and is hosted by the search engine company [7]. *Web search queries* are expressions in the form of words, short sentences, or questions made by the user in the search engine to locate information [7]. Web search queries can be categorized as follows [7]:

*Informational queries:* These are web search queries in which the user is concerned with searching for information on general topics, e.g. "Cars" and "Airplanes".

*Navigational queries:* These are web search queries in which the user is concerned with searching for a specific website e.g. "Facebook".

*Transactional queries:* These are web search queries in which the user is concerned about searching an item to make a transaction such as purchase books, etc. "Buy books". *K-anonymity* is a data privacy technique proposed by Samarati and Sweeny that employs both generalization and suppression of values by requiring that for a database with sensitive values in an attribute, those sensitive values in that attribute be repeated at least *k>1* times to guarantee confidentiality [8].

*Permutation of web search query types* is a web search privacy heuristic outlined by Mivule (2017) in which a set of queries based on the different query categories, are combined to formulate a single query that is then executed in a search engine [9]. Mivule notes that typically, four major query categories are considered, namely, the *Navigational* queries denoted by *N*. *Informational* queries denoted by *I*. *Transactional* queries denoted by *T*. *Natural language processing* queries denoted by *L*. *Temporal* queries denoted as *P*. Furthermore, Mivule notes that a set of dummy queries is then generated and combined with the original query such that the proposed set of the search queries will be a variation of values in set *Q = {NITLP}*. This necessitates that any query set will include a permutation of *N, I, T, L, P* query phrases in combination with the original query. Additionally Mivule notes that the number of permutations of any *k* values can be computed formally in the following way [9]:

$$P(n,k) = \frac{n!}{(n-k)!} \; for \; 0 \leq k \leq n \tag{1}$$

In this case *n* represents the quantity in set *Q*, and *k* is the number of variations for any *k* values. Supposing the set *Q = {NITLP}*, the number of variations of the five items in the set *Q* is:

$$P(n,k) = \frac{n!}{(n-n)!} = \frac{n!}{0!} = n! \tag{2}$$

*If P*(5, 5), *then* 5 * 4 * 3 * 2 * 1 *will result in* 120.

Therefore the number of rearrangements of the five items in set *Q* results in 120 permutations of possible obfuscated query arrangements [9]. *Semantic search* is an aspect of web search queries that focuses on meaning, understanding, intimation, and inference of a web search query [7].

*Text Mining*: To measure the effectiveness of the search query obfuscation methods, text mining is applied on the retrieved documents, in the form of snippets, to quantify the frequency of words that relate to the obfuscated search query keywords. Text mining then allows for the quantification of how many documents relate to dummy query keywords and how many relate to the actual query keyword being obfuscated. The basic observation at this point in the experiment, is that the more documents relate to dummy query keywords, the more distortion and thus obfuscation. A lower correlation corresponds to lower privacy. Consideration must be given regarding privacy vs. usability. The higher the distortion in the retrieved documents (dummy search query related documents) the less usability that results (less relevant documents), and vice versa. A user defined balance between privacy and usability requirements should always be sought considering the necessary trade-offs. The first step in the text mining process [10]-[12] is to read the retrieved documents from the search engine. For this study, documents were retrieved in the form of search engine snippets and were stored for each query that was executed. For each query, the goal was to retrieve 100 snippets – representative of the *top-k* results, in this case top-100 documents. The next step is to process text documents based on a TF-IDF word vector generation. *TF-IDF* is the Term Frequency – Inverse Document Frequency metric used to evaluate how important a word is in a corpus [13]. Several key steps follow, as described in the remainder of this subsection:

*Transform cases*: All characters in the text documents are subsequently transformed to lower cases or upper cases.

*Filter stop words*: The next phase of the text mining process is to filter stop words, which involves removing commonly used English words from the text documents, for example, "*this*", "*that*", and "*you*".

*Tokenization*: Tokenization is then applied to the text at this time, a process that separates each word in the text document as a single entity, which is later used to build a word vector.

*Stemming*: At this phase of the text mining process, stemming is a procedure in which words are reduced to their original root form. For example, we might reduce the word, "mining", to "mine".

*N-Grams*: This is performed along with stemming. N-Grams involve finding a group of words or terms that repeat together within a given length. For example, we might be interested in finding a group of length-two words, "*US Dollar*" that appear together in a financial blog.

*Machine Learning*: One of the motivations for using machine learning in this study is to classify web search queries into two categories, real and dummy queries. If the classification accuracy is high then the dummy queries are discernable. Otherwise, if the classification accuracy is low then dummy queries are indistinguishable, and this is likely to be an indication of better query obfuscation and privacy. Researchers have observed that KNN classification algorithm was very effective in the categorization of documents due to the availability of the similarity metric (Euclidean Distance) that can be employed in categorizing neighbors of a specific text [14]. Preliminary results from this study seem to corroborate this observation; KNN classification outperformed other classification algorithms that were used in this study. KNN utilizes the Euclidean Distance measure in categorizing items. The Euclidean distance measure can formally be noted as follows [15], [16]:

$$distance(x,y) = \sqrt{\sum_{i=1}^{n}(x_i - y_i)^2} \tag{3}$$

The KNN process using the Euclidean distance works by 1) computing the distance between item *d* in the training set and each example in *D* in the assessment set; 2) KNN then chooses the *k* examples in the assessment set *D* nearest to those in the training set *d*, and labels the set by *P* a subsection of *D*; and 3) KNN then assigns training set item *d* to the most mutual class in *P*.

*Precision and recall* are the two key metrics used by search engines to analyze the effectiveness of web search queries in the context of relevant documents out of all the retrieved documents as a result of a search query execution. The rate for these metrics lies between 0 and 1, whereby 1 is the highest value indicating ideal return for precision and recall [9]. Precision and recall can be expressed mathematically as follows:

$$Precision\ (P) = \frac{Total\ of\ retrieved\ relevant\ articles}{Total\ of\ retrieved\ articles} \quad (4)$$

$$Recall\ (R) = \frac{Total\ of\ retrieved\ relevant\ articles}{Total\ of\ relevant\ articles} \quad (5)$$

B. Related works

*Obfuscation of web navigation profiles*: Dankar and El Emam, (2013) proposed a web search privacy technique for the obfuscation of web navigation profiles that involved using dummy web search queries generated by $k$ different user profiles to make web search requests equivalent to those of the real user. Here, the parameter $k$ is adjustable to the number of obfuscation profiles to satisfy a user's privacy needs [17]. Our suggested heuristic differs in that we primarily use a permutation of web search queries and do not depend on multiple users to generate a user web search profile. In our case, the user profile is generated by the strategic k-anonymity clicks on the document URLs and Ads generated by the search results.

*Plausible deniability*: Plausible deniability search is a web search query privacy procedure in which a set of $k-1$ dummy queries with traits similar to the original, but on unrelated topics, are generated and used to conceal and disguise original queries [9], [18]. Plausible deniability search dictates that every original query be substituted with a regular but analogous dummy query intended to retrieve results very similar to those expected from the original query. Any subset of $k$ dummy queries will produce statistically indistinguishable results to an equivalent original set of $k$ queries [9], [18]. The generated dummy queries are executed at the same time to conceal the intent of the user, making it challenging to detect which specific query was intentioned by the user; leaving the burden of proof to the search engine to establish which query belonged to the user [9], [18]. While we meet some criteria for plausible deniability search, our heuristic approach digresses by generating dummy queries based on query type rather than query topic. In our case, we are interested in generating dummy query keywords that include a permutation of navigational, informational, transactional, temporal, and natural language processing query keywords. The query keywords can be related or unrelated to the original query. In either case, they are combined together with the original query keywords during query execution. Each permutation is expected to produce its own varying results.

*Private, efficient and accurate web search (PEAS)*: PEAS is a search privacy solution proposed by Petit et al (2015) to prevent information leakages as a result of numerous web search queries issued by users on search engines that link users to sensitive information. Petit et al. (2015) proposed a solution that includes a mechanism that hides users' identities by breaking any links between queries and users, and an obfuscation mechanism that injects noisy queries among the real queries to provide privacy [19].

*Dummy query generation*: In their evaluation of dummy query generation methods, Balsa et al. (2012) observed that while dummy queries prevent the precise deduction of search profiles, by providing query deniability, such obfuscation also generally reduces the usefulness of search profiles [20]. Balsa et al. also noted that canonical queries used in the obfuscation process decrease the usefulness of the search outcome since they are not precise. Balsa et al. suggested that effective obfuscated queries should guarantee that actual and dummy queries act identically in terms of their usefulness [20].

*Search embellishment:* Search embellishment is a model created to implement user intent privacy with improved retrieved result usability. Each query search keyword is embellished with distraction keywords that are similar to the actual user search terms with plausible deniability traits, but pointing to different plausible topics [21]. The distraction keywords are infused directly into each user search query to prevent the search engine from deciphering user intent based on query search keywords [21]. Pang et al. apply only a small subclass of search terms from a corpus to serve as decoys and reduce the high computational overheads associated with the use of the whole corpus. To avoid inference attacks, Pang et al avoid using semantically related terms that point to the same topic, and avoid placing together search terms that are highly specific and frequent [21]. In our suggested heuristic, rather than avoiding highly specific or frequent search terms as in the Pang et al model, we capitalize on permutations of query type search terms that could be highly specific, less specific, frequent, or infrequent but that provide distortion of the query results, making it difficult for the search engine to pinpoint specific user intent. Researchers have observed that while dummy search queries prevent the precise inference of user search profiles, such obfuscation always reduces the utility of search profiles and retrieved results, while efficient obfuscated queries should guarantee that actual and dummy queries are indistinguishable [20]. Our heuristic model diverges from this standard approach by capitalizing on noise and the distortion of retrieved search query results. Control is given to the user to click $k>1$ diversionary and actual URLs to obfuscate user intention, making it difficult for the search engine to decipher specific user intent.

*Topical intent obfuscation*: To obfuscate user intent, Wang and Ravishankar (2014) proposed using a combination of themes in the real queries that match those of the dummy queries. The selected keywords in the dummy queries are derived from the same theme as those from the original queries, making it problematic to semantically differentiate dummy queries [22]. Moreover, Wang and Ravishankar (2014) observed that achieving an entirely secure keyword-based obfuscation technique is intractable, since a secure technique must warrant that real and dummy queries are the same and indivisible [22]. However, the Wang and Ravishankar methodology would be vulnerable to inference attacks via generalization. For instance, if all search queries formulated in a time period $t_1$ to $t_n$ are indistinguishable because of a strong semantic relation, an attacker with sufficient resources might only have to look at the general theme and analyze the series of query requests to predict specific user intent [23], [24]. To avoid the possibility of this type of attack on specific user intent privacy, our heuristic capitalizes on using distortions during query formation by avoiding keywords that are highly semantically related. Dummy keywords in our suggested query formation could include unrelated topics, so as to provide enough specific user intent concealment.

*TrackMeNot* is a web search query obfuscation technique that utilizes a web browser to execute dummy web search queries in conjunction with the original queries by randomly generating the dummy search query keywords from RSS news feeds. User control of their own intent privacy is controlled in the choice of the dummy search queries in that TrackMeNot permits the user to choose RSS feeds based on a particular topic but the random selection of the dummy search query keywords is left to the TrackMeNot algorithm [25], [26]. Our proposed Distortion Search heuristic differs from TrackMeNot, in that we do not utilize RSS feeds to generate dummy queries and our approach is not browser-based. Rather, Distortion Search relies on a systematic generation of dummy search queries using the permutation of web search query types.

III. METHODOLOGY

The term "*distortion*" is used to imply distorted retrieved search results, a consequence of actual keywords and highly visible dummy keywords in the obfuscated search query. The aim at this point is not to do away with distorted retrieved results, but rather to use such noisy retrieved search results as an obfuscation cover for specific user intent and the generation of a user pseudo-profile.

*A. Distortion Search Hueristic*

**PROCEDURE 1:** Distortion Search Heuristic
1. Permutation and search query generation
    a. Selecting highly visible search engine keywords
    b. Selecting semantically related verb keywords
    c. Selecting search query permutations
    d. Combine original query with query permutations
    e. Executing new generated search query permutations
2. *K>1* strategic URL clicks.
3. Search results retrieval.
4. Text mining retrieved search result.
5. Machine learning to predict actual queries vs. obfuscated queries.
6. Tracking Ads.
7. Overall obfuscated query performance.

The methodology process for the distortion search heuristic is as follows:

1) *Obfuscated query formation*: In this process, permutation techniques are applied to the search query to generate various arrangements of keyword types for specific user intent obfuscation as shown in Fig. 3 and 4. Likewise highly visible search engine phrases and verb synonyms are generated and used for intent obfuscation.

2) *Text-mining*: The next phase involves text mining as shown in Fig. 2. In this stage retrieved web search query results, snippets in this case, are text-mined and information retrieval metrics, such as recall, average precision, and f-measure, are used to measure search query performance. In this case, we are interested in documents related to the real search key phrase, "Buy Toyota" in the top-k retrieved results.

3) *Machine learning query prediction*: In this phase, machine learning methods are used to determine if obfuscated queries can be predicted from real queries. In this case, real AOL queries released in 2006 [27] are used to compare with obfuscated queries generated from the distortion search heuristic. The more obfuscated queries are in similarity to real queries, the more likely that privacy levels are low in terms of the retrieved documents. In this way, finding the right balance between privacy and usability remains a challenge.

4) *Overall performance*: In this phase of the experiment, the overall performance of the obfuscated search queries is analyzed and recommendations are made for refining or fine-tuning the distortion search obfuscation techniques.

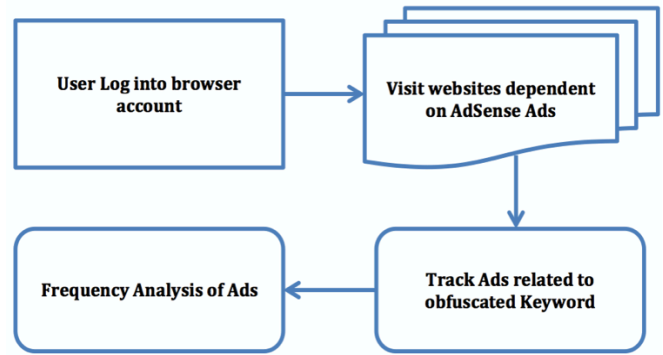

Fig. 1. Ads tracking process.

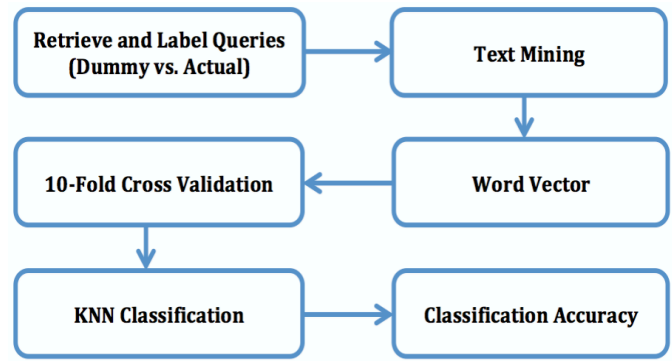

Fig. 2. Text mining process.

*B. Obfuscated Query Formation*

The following are the steps taken in the formation of obfuscated search queries:

1) The first step is to identify the main intended original keyword or search term. This is the keyword or search query term to obfuscate.

2) The second step is to identify verbs related to the main search term or keyword. Verbs are good at revealing specific user intent.

3) The third step is to generate semantically related verb terms to include in the obfuscated query. Similar verbs that are one or two degrees separated from the root verb would be ideal.

4) The fourth step is to identify highly visible dummy keywords or search terms that might or might not be semantically related to the main intended original keyword, but would return highly relevant retrieved search results.

5) The fifth step is to categorize the main intended original keyword, and dummy keywords into query type classes – informational, navigational, transactional, temporal, and natural language processing query types.

**6)** The sixth step is to generate permutations of the query types. For example, informational and temporal query, transactional and natural language processing query type, etc.

7) The seventh step is to execute the search queries derived from the permutation of query types at the same time, as a single obfuscated query. For example, the query {Honda, 2014, Barack Obama} derived as a permutation of informational and temporal search keywords is executed as a single obfuscated query.

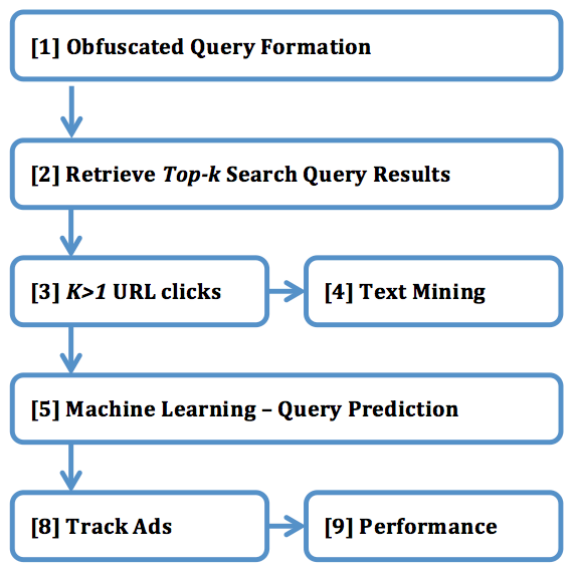

Fig. 3. An overview of the distortion search process.

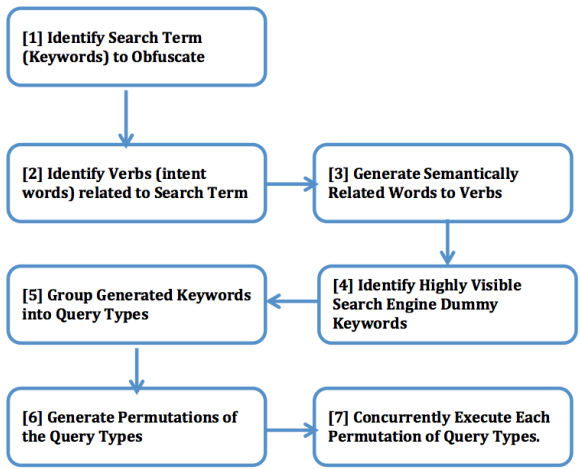

Fig. 4. Permutation process.

## C. Experiment

The objective of the testing in this study was to 1) privatize queries using the distortion search heuristic; and 2) privatize user intent in regards to the search queries. The experiment involves a user whose true intent is to search the web and buy a Toyota car [9]. More particularly, the user wants to execute the following query privately, *"buy a Toyota"* car but without revealing their intent to the search engine. The experiment involved the following steps [9]:

1) In the experiment, embedding the original query in a set of diversionary keywords to obfuscate the actual search query was done. The retrieved outcomes ought to, for example, to permit the user to browse Toyota related snippets or webpages while clicking on other presented web search snippets as a diversionary scheme.

2) The intention of the experimentation was to apply the distortion search heuristic to disguise the search query, "Buy Toyota", and specific intent that the user premeditated to "buy a Toyota".

3) The ensuing permutations were chosen for the initial experiment: Q ={I, IT, IP, TP, IL, NI, NIT, NIP, IPL, ITP, NITP, ITPL, NIPL, NITL, NITPL}. This means that a query could contain only informational search keywords, contain both informational and transactional keywords, contain both informational and Natural Language processing keywords, and so forth.

4) An overall 121 search queries were produced (see Fig. 5) after the permutation procedure that generated about 12,000 retrieved Google documents for the study. Each produced query permutation was then executed simultaneously with the original query.

5) During this procedure, the user is logged into their Google-Chrome search engine/browser account to keep a log of the user search and browsing history.

6) An overall of 12,177 Google snippets (text documents) were retrieved. Analysis was then done using text mining tools to differentiate relevant from non-relevant documents.

7) Snippets that contained the terms "Buy Toyota" were considered relevant.

8) Machine learning was then applied on the search queries to determine if obfuscated queries could be separated from real queries. A publicly available AOL query dataset released in 2006 was used as a benchmark to determine if obfuscated queries could be predicted from actual queries.

9) Due to privacy and legal concerns, many entities won't publish search query logs and as such, the 2006 AOL data set had to be adopted for this experiment.

10) The user then strategically clicked on k>1 URLs generated by both dummy queries and the original query. In this case, the user would click on k>1 URLs generated by the real "Buy Toyota" links and the k>1 URLs generated by the other dummy queries. Furthermore, the user clicked on k>1 AdSense links. The goal was to generate a random and noisy browsing history, making it difficult for to track user intent.

11) The last part of the experiment was to track AdSense ads generated as a result of the user search and browsing history.

12) The user specifically and intentionally visited a total of 16 websites and blogs for a period of seven days. Only AdSense Ads were tracked on each homepage of the visited website as illustrated in Fig. 1. Empirical results were then generated and analyzed.

13) The assumption in this experiment is that the search engine does not have apriori knowledge as to what the original user intent is in formulating their search queries.

| Sample of Web Search Query Permutations | | |
|---|---|---|
| Q82 | NITP | Nairobi kenya, influencence, shoes.com, get a samsung phone, buy a toyota 2014 |
| Q111 | IT | Aquire books |
| Q81 | NITPL | I want a toyota 2014, get a samsung phone, shoes.com, nairobi kenya, influence |
| Q80 | NITP | Buy a toyota 2014, get a samsung phone, shoes.com, Nairobi Kenya, influence |
| Q12 | NITP | Acquire Toyota 2014 get Samsung Obtain shoes cnn.com western civilization |
| Q28 | NITP | Nairobi kenya, influencence, shoes.com, get a samsung phone, buy a toyota 2014 |
| Q110 | I | Influence Coffee |
| Q2 | NI | Honda Car Kamplalal Blue Jays Cnn Forecaster Franc motorolaToyota recall precision |
| Q4 | NIT | Honda Green Blue Sell 2011 Peyton Manning Toyota Kampala Purchase When  Get Car |
| Q108 | IT | Get (Honda) |
| Q1 | NI | bbc honda afric newton blue jays freetoyota r us peytonmanning blueribbons |

Fig. 5. A sample of web search query permutations.

## IV. RESULTS

The first phase of the preliminary results analysis looks at how many retrieved documents were relevant based on the results of the query permutations [9]. In Fig. 6, retrieval search results from the obfuscated queries are presented with a total of 121 queries. The number of retrieved and relevant documents, snippets in this case, is presented with queries Q1 to Q121 on the x-axis and quantity on the y-axis respectively. The key web search query term "Toyota" was used in determining all articles that were considered relevant in this experiment.

Fig. 6 shows results from the 121 query permutations that were executed. The x-axis represents the queries from Q1 to Q121, while the y-axis represents that number of items retrieved. The blue shade in the graph represents the overall documents retrieved, while the red shading represents the relevant documents that were retrieved. For example, we see in Fig. 6 that query Q17, returned a total of 106 documents yet only 53 of those 106 retrieved documents were relevant. In this case only 53 out of the 106 documents contained the phrase "Toyota" to be counted as relevant. The rest of the 53 retrieved documents that are not relevant are representative of the diversionary keywords and phrases added to obscure the original query and user intent. In other words, the remaining 53 documents that are not relevant are used as a decoy to hide the original user intent. In this case, the intent of the user to "Buy a Toyota" remains obscured and hidden with the retrieval of non-relevant documents. The number of articles retrieved was about 100 articles for every query executed; only the *top-k* documents were chosen, were $k$ is about 100 documents on average [9]. Studies have indicated that on average users tend to scan only the first page of the search engine results scanning through the first 20 snippets on the first present page of search results [9], [28], [29]. It is important to note that queries that returned a high number of relevant documents are those in which the original search key phrase, "Buy Toyota", is not so much obscured. In other words, query permutations with few extra diversionary keywords returned higher number of relevant documents while those with more diversionary keywords in the permutation returned less relevant documents. This outcome was expected as it has been well documented in literature about the tension between privacy and usability. The more privacy we had injected in the query permutation (more diversionary keywords), the less usability (less relevant documents), and vice versa. Finding the balance between privacy and usability needs, remains a challenge as this study further shows, and as such, requiring tradeoffs [30]-[35].

*A. Machine Learning – Obfuscated vs. AOL Queries*

This section presents experimental results from applying machine learning to classify whether queries are obfuscated or real. One of the motivations to using machine learning was to determine if obfuscated queries could be predicted from real queries. If prediction were difficult then it would indicate that the obfuscated query could withstand a classification attack from a search engine with enormous computation resources. Results from this phase of the experiment could be used to refine the query obfuscation process to generate obfuscated queries that are, ideally, indistinguishable from actual user queries.

Two parts of this experiment were implemented. In the first part, machine-learning algorithms were applied to classify dummy queries generated by TrackMeNot and actual queries from the AOL dataset published in 2006. TrackMeNot was used as a benchmark to test our hypothesis. A sample of 1,392 queries was used in the TrackMeNot vs. AOL experiment. 390 dummy queries were taken from TrackMeNot. The remainder consisted of real queries from the AOL dataset. A total of 370 queries were analyzed in the Distortion Search vs. AOL experiment, with 122 belonging to distortion search and 248 consisting of AOL queries. A binary labeling method was used for the dataset, 0 for dummy TrackMeNot queries and 1 for real AOL queries. This followed after the process of text mining and extracting the word vector. A 10-fold cross-validation was used in the classification process. The classification accuracy in predicting dummy TrackMeNot from real AOL queries registered at 80.45, 84.26, 71.76, and 72.11 for KNN, Naïve Bayes, Random Forest, and Logistic Regression, respectively, as illustrated in Fig. 7. The classification accuracy in predicting distortion search obfuscated queries from real AOL queries registered at 78.31, 77.57, 67.3, and 52.16 for KNN, Naïve Bayes, Random Forest, and Logistic Regression, respectively. The obfuscated queries generated using the distortion search heuristic performed better with lower classification accuracy compared to the TrackMeNot vs. AOL queries. For instance, KNN returned a classification accuracy of 80.45 for TrackMeNot vs. AOL compared to 78.31 for Distortion Search vs. AOL. This could be interpreted as better obfuscation for distortion search. Lower classification accuracy is desired, as this would indicate that, at some level, the obfuscated queries are indistinguishable from the actual queries. Results from this part of the experiment could provide the privacy curators with insight into how to adjust the distortion search heuristic to generate obfuscated queries that are indistinguishable from real queries. However, it is important to note that much lower classification accuracy results might be interpreted as being generated by obfuscated queries that in some sense are similar in traits to the original queries and as such not private enough. On the other hand, highly distinctive search queries and keywords might help with greater distortion and decoy effects, and could therefore help in improving obfuscation measures, while dummy queries that are very similar to the real queries might not. Therefore the notion of privacy vs. usability comes into play once again in such scenarios and, as such, privacy curators must take into account the trade-offs that need to be made.

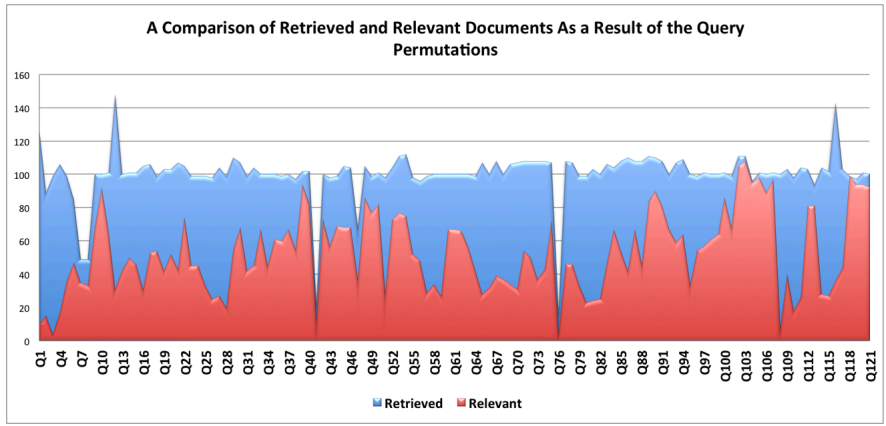

*Fig. 6.* Retrieved vs. Relevant documents per search query.

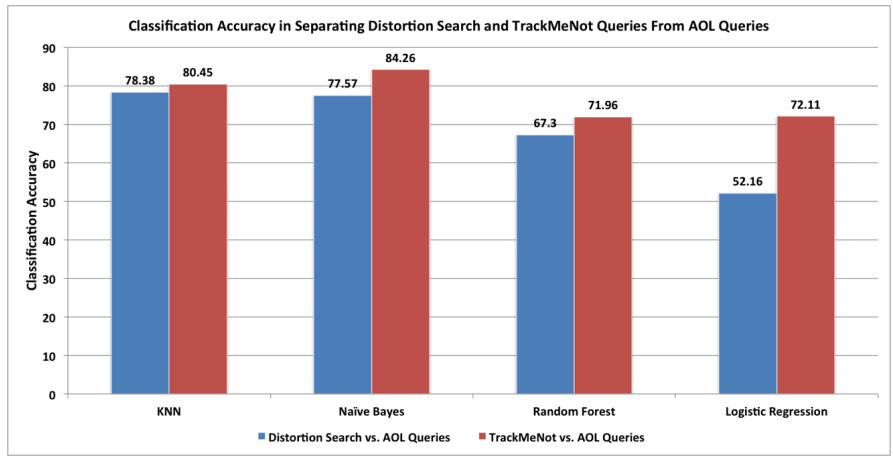

*Fig. 7.* Classification accuracy in predicting obfuscated queries from real queries.

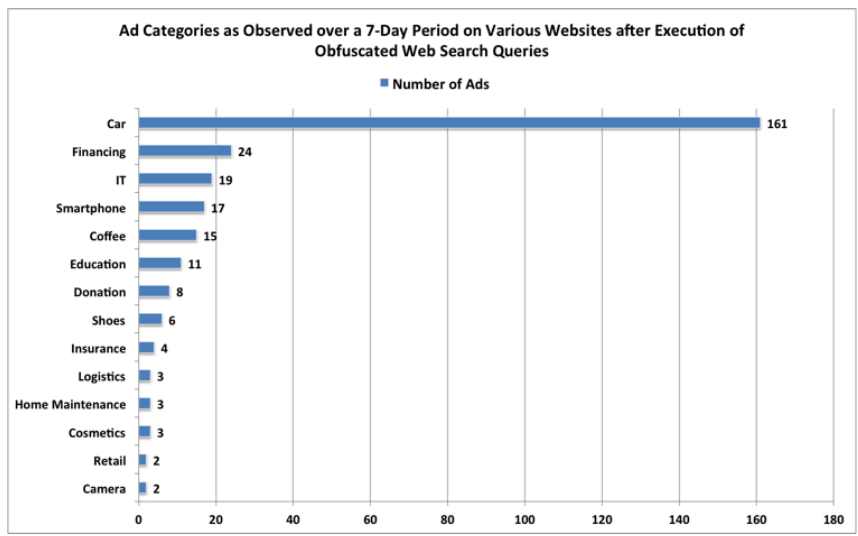

(a)

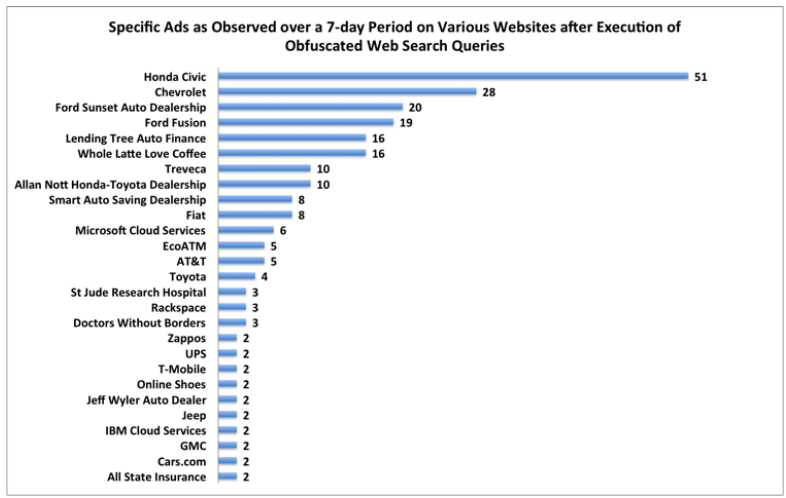

(b)

Fig. 8. (a) Number of ads in the conceptual categories (b) number of ads in the specific user intent categories.

## B. Tracking Ads

The last phase of the experiment dealt with tracking AdSense ads generated as a result of the user search and browsing history. The user visited a total of 16 websites and blogs for a period of seven days while logged into their Google account as illustrated in Fig. 1. The motivation was to use Google's generated browsing and search history storage capabilities that in turn link directly to the AdSense ads. This ensures that the ads generated are based solely on the user's historical browsing activity. Only Google AdSense ads were tracked on each homepage of the visited websites. Empirical results were then generated and analyzed.

Fig. 8(a) and (b) show the number of ads generated in each conceptual and specific user intent category. While the overall goal was to obfuscate specific user intent, an analysis of the conceptual user intent is vital in the generation of future fine-tuned obfuscation techniques. In this case, the conceptual user intent keyword would be "*cars*", while the specific user intent keyword would be "*Toyota*". One of the goals of this study was to obfuscate specific user intent; in this example, the search engine could know that the user was generally searching for "*cars*" but not the specific car that the user intended to buy, in this case, a "*Toyota*". Out of a total of 293 ads, 161 ads were listed under the "*Cars*" category, as illustrated in Fig. 8(a). The remaining 132 ads were distributed among the various categories including financing, Information Technology (IT) and Smartphones, in particular. These results suggest that the search engine is able to discern the conceptual intent of the user – in this case the user probably is searching for a "*car*" or needed "*financing*", which yields ads in financing category.

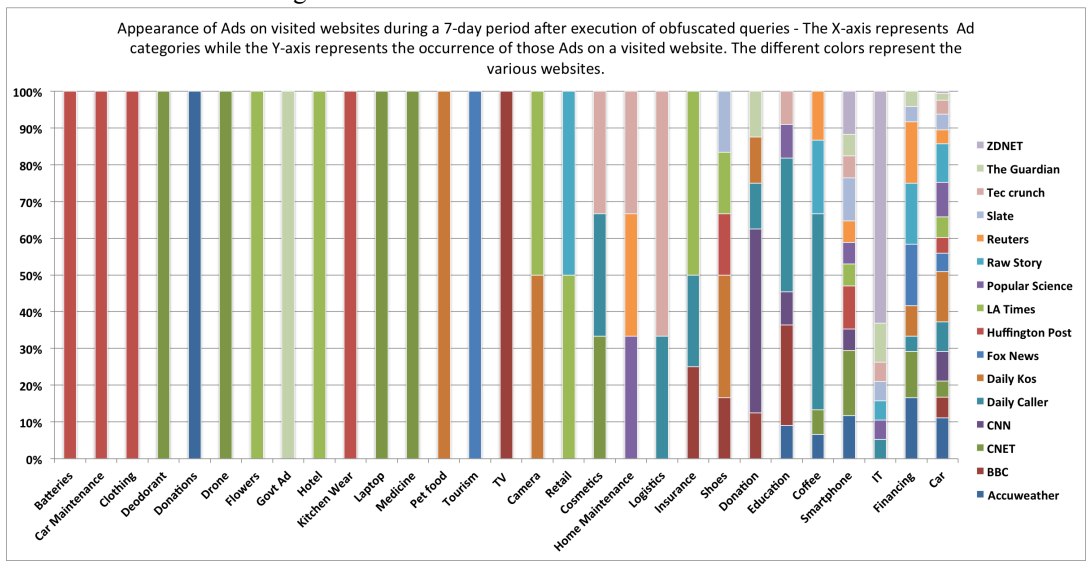

Fig. 9. Category Ad appearance on visited websites during 7-day period.

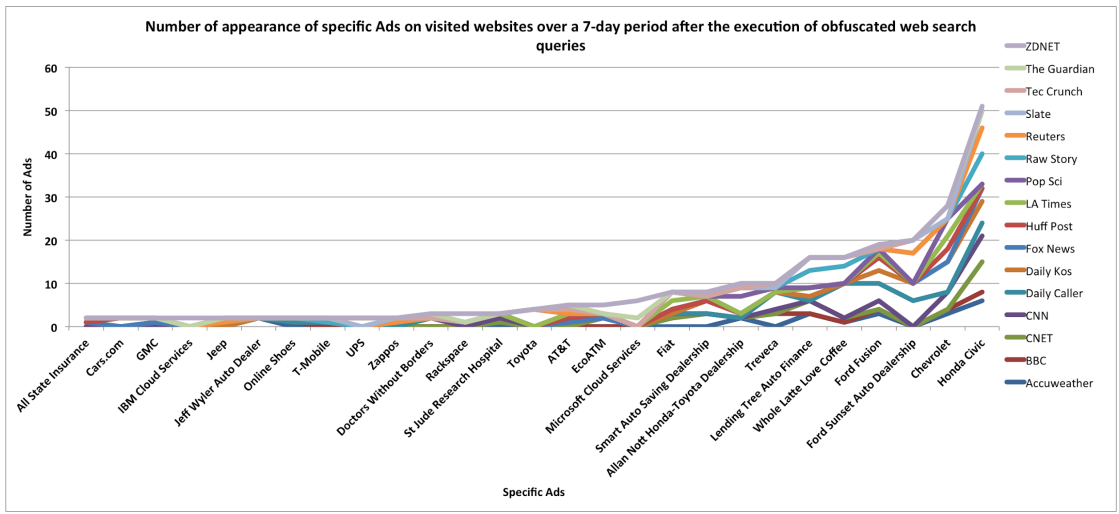

Fig. 10. Specific ad appearance on visited websites during 7-day period.

However, the search engine supposes that the user is also searching for IT and Smartphones – categories that relate to the dummy search terms used in the obfuscated query during the distortion search process. These results could assist the privacy curator in adjusting the distortion search heuristic to, for example, increase the number of IT and Smartphone related search key terms while reducing car related search keywords but with the privacy vs. usability trade-offs in context. In terms of the specific user intent category, Fig. 8(b) shows the results of specific ads targeted at the user. A total of 293 ads were observed over a 7-day period visiting 16 websites and blogs that depend on AdSense ads for income generation. Out of the 293 ads, only 14 ads were specifically related to "*Toyota*" or "*Buy Toyota 2014*". In this case, only 7.142% of the ads addressed the user's specific intent, and that was to "*Buy Toyota 2014*". 93 percent of the remaining ads were not directly related to the specific user intent, as illustrated in Fig. 9 and 10. This is further illustrated in Table 1 (Appendix A), which shows that there were only four specific *Toyota* ads observed on the *Pop Sci* website. The other 10 ads related to *Toyota* had to do with financing and auto dealerships. The results suggest that the distortion search heuristic was successful in obfuscating and providing privacy for specific user intent. In this web search experiment, specific user intent was successfully obfuscated 93 percent of the time.

## V. CONCLUSION AND FUTURE WORK

Privacy remains a major concern when making use of search engines to retrieve information on the Internet. Search engines maintain enormous computational capabilities to keep Internet search logs for each user, which presents major privacy worries for both individuals and organizations that depend on the search engines for information retrieval. While a number of web search query privacy heuristics have been suggested, many approaches require collaboration with search engines and other third party entities, making web search query privacy a major challenge when trying to hide intent from search engine providers. The main contribution in this article is the presentation of Distortion Search, web search query privacy heuristic that works by 1) the formation of obfuscated search queries via a permutation of query keyword categories; and 2) the application of k-anonymised web navigational clicks to generate a distorted user profile and thus provide user intent and query confidentiality without the need of search engine or third party collaboration. The article presents empirical results from the evaluation of distortion search queries in terms of retrieved and relevant search results and AdSense web ads. Preliminary results from this study show that is might be possible to provide confidentiality to user intent when executing web search queries. Additionally, the issue of privacy vs. usability still haunts web search query privacy. Trade-off between data privacy vs. usability remains a difficult problem. Too much obfuscation could return less relevant search results, while less obfuscation could return highly relevant retrieved search results but at the cost of revealing user intent and of weakened web search confidentiality. Additional research is necessary in the creation of confidential web search queries that take into consideration both privacy and usability questions. Additionally, the issue of search engines and their computational capabilities should be a major concern to all privacy custodians. With machine learning algorithms and high performance computing, data privacy heuristics such as distortion search are vulnerable to classification attacks and therefore such privacy algorithms need constant revision. Future work will involve the large-scale study and application of the distortion search heuristic on bigger datasets operating on enterprise systems with multiple logged in users. A consideration of classification attacks on Distortion Search and how to thwart such attacks will be done. A much more detailed comparative study of the Distortion Search in regards to different query types will be done. While search queries in this preliminary study centered around the search terms "buy Toyota", we intend to fully expand on different search terms from the informational, navigational, transactional, and natural language processing query types. This will include an investigation of which search key phrase permutations work best in obfuscating user intent while allowing for a fair amount of relevant retrieved results.

APPENDIX A

| Ad | Accuweather | BBC | CNET | CNN | Daily Caller | Daily Kos | Fox News | Huff Post | LA Times | Pop Sci | Raw Story | Reuters | Slate | Tec Crunch | The Guardian | ZDNET | Grand Total |
|---|---|---|---|---|---|---|---|---|---|---|---|---|---|---|---|---|---|
| All State Insurance | | | | | 1 | | | 1 | | | | | | | | | 2 |
| Cars.com | | | | | | | | 2 | | | | | | | | | 2 |
| GMC | | | | 1 | | | | 1 | | | | | | | | | 2 |
| IBM Cloud Services | | | | | | | | | | | | | | | | 2 | 2 |
| Jeep | | | | | | | 1 | | | | | 1 | | | | | 2 |
| Jeff Wyler Auto Dealer | 2 | | | | | | | | | | | | | | | | 2 |
| Online Shoes | | 1 | | | 1 | | | | | | | | | | | | 2 |
| T-Mobile | | | 1 | | | | | | | | | 1 | | | | | 2 |
| UPS | | | | | | | | | | | | | 2 | | | | 2 |
| Zappos | | | | | 1 | | | | | | | 1 | | | | | 2 |
| Doctors Without Borders | | | 2 | | | | | | | | | | | | 1 | | 3 |
| Rackspace | | | | 1 | | | | | | | | | | | | 2 | 3 |
| St Jude Research Hospital | 1 | | 1 | 1 | | | | | | | | | | | | | 3 |
| Toyota | | | | | | | | | | 4 | | | | | | | 4 |
| AT&T | | | 1 | | | | 1 | 1 | | | | 1 | | | 1 | | 5 |
| EcoATM | | | 2 | | | | 1 | | | | | | | | | 2 | 5 |
| Microsoft Cloud Services | | | | | | | | | | | | | | 2 | | 4 | 6 |
| Fiat | | 2 | 1 | | | 1 | | | 2 | 2 | | | | | | | 8 |
| Smart Auto Saving Dealership | | 3 | | | | 3 | | 1 | | | | | | | 1 | | 8 |
| Allan Nott Honda-Toyota Dealership | 2 | | | | 1 | | | | | 4 | 2 | | | | 1 | | 10 |
| Treveca | | 3 | | 1 | 4 | | | | | 1 | | | | 1 | | | 10 |
| Lending Tree Auto Finance | 3 | | 3 | | | 1 | 2 | | | | 4 | 3 | | | | | 16 |
| Whole Latte Love Coffee | 1 | | 1 | | 8 | | | | | | 4 | 2 | | | | | 16 |
| Ford Fusion | 3 | 1 | | 2 | 4 | 3 | 3 | | 1 | 1 | | | | | 1 | | 19 |
| Ford Sunset Auto Dealership | | | | | 6 | 4 | | | | | 7 | | 3 | | | | 20 |
| Chevrolet | 3 | | 1 | 4 | | 7 | | 3 | 3 | 4 | | | 3 | | | | 28 |
| Honda Civic | 6 | 2 | 7 | 6 | 3 | 5 | 3 | | 1 | | 7 | 6 | 4 | | | 1 | 51 |
| TOTAL | 20 | 14 | 14 | 18 | 29 | 26 | 10 | 8 | 10 | 16 | 24 | 12 | 10 | 6 | 7 | 11 | 235 |

Fig 11. NUMBER OF SPECIFIC ADS THAT APPEAR MORE THAN ONCE AND THE RELATED VISITED WEBSITE OR BLOG